\documentclass{article}
\usepackage{spconf,amsmath,graphicx,hyperref}
\usepackage{amssymb}  
\usepackage{amsfonts}
\usepackage{svg}
\usepackage{url}
\usepackage{booktabs}
\usepackage{enumitem}
\usepackage{subfigure}
\usepackage{booktabs,multirow,pifont}
\usepackage{textcomp} 
\usepackage{cite}  
\usepackage{threeparttable} 
\newcommand{\cmark}{\ding{51}}
\newcommand{\xmark}{\ding{55}}

\newcommand{\sys}{DMP-TTS}
\title{\sys: Disentangled multi-modal Prompting for Controllable Text-to-Speech with Chained Guidance}
\name{
\begin{tabular}{c}
Kang Yin $^{1,\dagger}$,
Chunyu Qiang$^{2,\dagger}$,
Sirui Zhao $^{1, *}$, 
Xiaopeng Wang$^{2}$, \\
Yuzhe Liang$^{2}$, 
Pengfei Cai$^{1}$, 
Tong Xu$^{1,*}$, 
Chen Zhang$^{2}$, 
Enhong Chen$^{1}$
\end{tabular}
}

\address{
$^1$ University of Science and Technology of China, Hefei, China
$^2$ Kuaishou Technology, Beijing, China\\
}
\begin{document}
\ninept
\maketitle

\begingroup
  \renewcommand\thefootnote{}
  \footnotetext{$^\dagger$ denotes equal contribution. \textsuperscript{*} denotes corresponding author.}
\endgroup

\begin{abstract}
Controllable text-to-speech (TTS) systems face significant challenges in achieving independent manipulation of speaker timbre and speaking style, often suffering from entanglement between these attributes. We present \sys, a latent Diffusion Transformer (DiT) framework with explicit disentanglement and multi-modal prompting. A CLAP-based style encoder (Style-CLAP) aligns cues from reference audio and descriptive text in a shared space and is trained with contrastive learning plus multi-task supervision on style attributes. 
For fine-grained control during inference, we introduce chained classifier-free guidance (cCFG) trained with hierarchical condition dropout, enabling independent adjustment of content, timbre, and style guidance strengths.
Additionally, we employ Representation Alignment (REPA) to distill acoustic-semantic features from a pretrained Whisper model into intermediate DiT representations, stabilizing training and accelerating convergence.
Experiments show that DMP-TTS delivers stronger style controllability than open-source baselines while maintaining competitive intelligibility and naturalness. 
Code and demos will be available at \url{https://y61329697.github.io/DMP-TTS/}.

\end{abstract}
\section{Introduction}
\label{sec:intro}


TTS systems convert text into natural and intelligible speech. With recent advances approaching human-level quality \cite{ren2020fastspeech,kim2021vits,wang2023neural,qiang2024minimally,wang2025m3,qiang2025secousticodec,qiang2025vq}, research focus has shifted toward \emph{controllable} TTS\cite{jiang2024mega, peng2024voicecraft, kim2021expressive, gao2025emo, guo2023prompttts, yang2024instructtts, qiang2025instructaudio}, which seeks independent and flexible manipulation of diverse speech attributes for more natural human–computer interaction.
Existing approaches to TTS control mainly fall into three categories: reference-audio \cite{jiang2024mega, peng2024voicecraft}, discrete labels \cite{kim2021expressive, gao2025emo}, and descriptive text \cite{guo2023prompttts, yang2024instructtts, qiang2025instructaudio}. While these methods improve naturalness and controllability, they often entangle style and timbre, hindering independent control of attributes. 
For example, when using a reference audio for style control, timbre information from the reference often leaks into the generation, causing the synthesized speech to change timbre rather than preserving the intended target speaker \cite{xie2024towards}.
In addition, most mainstream TTS systems support only a single modality for style prompting (either audio or text), limiting their flexibility and adaptability.



Research into these challenges is at an early stage.
ControlSpeech \cite{ji-etal-2025-controlspeech} demonstrates multi-modal style prompting from both audio and descriptive text and takes a meaningful step toward style–timbre disentanglement. However, some practical considerations remain. First, the method is closely integrated with the NaturalSpeech3 backbone \cite{ju2024naturalspeech} for both architectural design and the disentanglement procedure, limiting extensibility and transfer to other backbones and deployment scenarios. Second, its textual prompts may include identity-related cues (e.g., gender), so the style channel can carry timbre information , complicating strict style–timbre separation and speaker consistency.



\begin{figure*}[!t]
    \centering
    \includegraphics[width=0.95\textwidth]{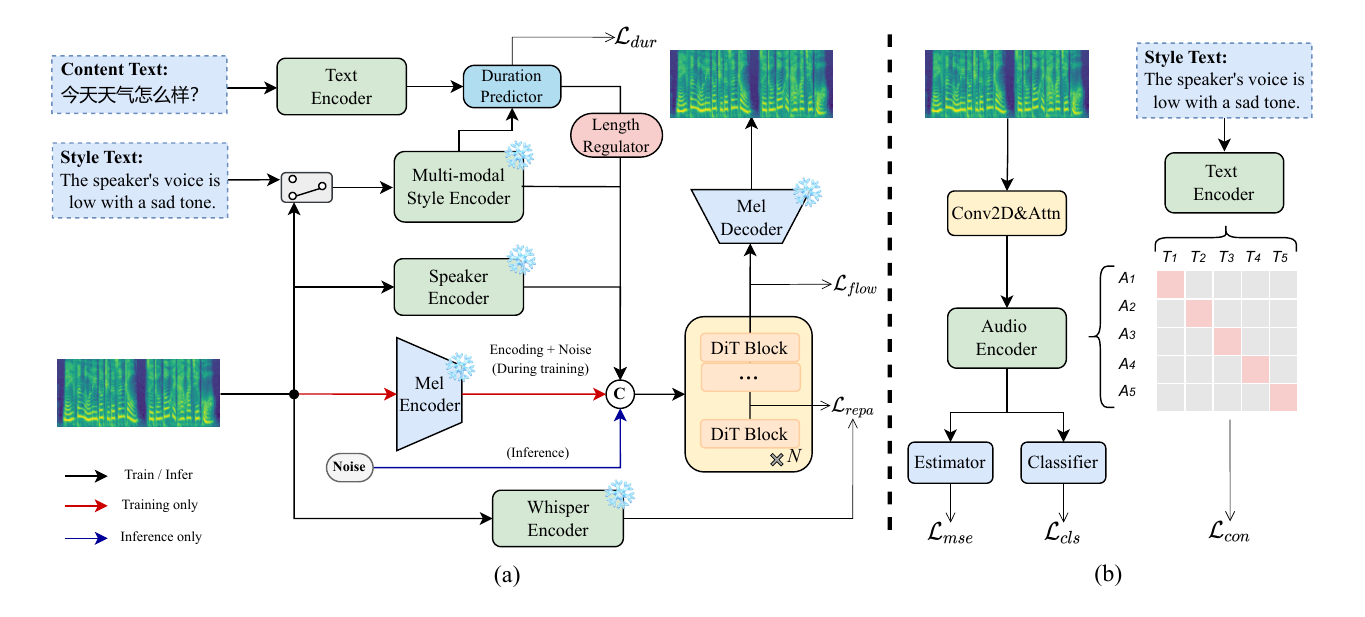}
    \vspace{-1.5em}
    \caption{(a) Overall architecture of \sys. (b) Unified multi-modal style encoder.}
    \vspace{-1.0\baselineskip}
    \label{fig:model_architecture}
\end{figure*}


To address these challenges, we present \sys, a flexible TTS framework built on a latent Diffusion Transformer (DiT) \cite{peebles2023scalable} with explicit disentanglement and multi-modal control. 
At its core, we introduce Style-CLAP, a unified multi-modal style encoder built on Contrastive Language–Audio Pretraining (CLAP) \cite{elizalde2023clap}, which maps style cues from reference audio and descriptive text into a shared embedding space.
We enhance style representation capacity with a multitask objective that predicts style attributes. 
For fine-grained, independent control at inference, we introduce chained classifier-free guidance (cCFG); trained with hierarchical condition dropout, it permits separate guidance strengths for content text, timbre, and style.
Finally, we improve convergence and stability via Representation Alignment \cite{yu2024representation} (REPA), aligning intermediate features with acoustic-semantic representations from a pretrained model. Our main contributions are:


\begin{itemize}[leftmargin=*, itemsep=0.3pt, topsep=0.4\baselineskip]
    \item We propose \sys, a DiT-based TTS framework that supports multi-modal prompting and achieves explicit style–timbre disentanglement.
    \item We build Style-CLAP, a cross-modal encoder that learns a shared, discriminative space for style attributes by aligning audio and descriptive text via contrastive learning and multi-task supervision.

    \item We introduce cCFG, a chained classifier-free guidance scheme that enables independent, continuous control of content, timbre, and style, and adopt representation alignment (REPA) to stabilize and accelerate training.
    
\end{itemize}


\section{Preliminaries}

We adopt Conditional Flow Matching (CFM) \cite{lipman2022flow} to train the latent DiT and briefly review it here.
CFM constructs a continuous-time flow between a noise sample $\mathbf{z}_1$ and a data sample $\mathbf{z}_0$ via a linear interpolation path:
\begin{equation}
    \mathbf{z}_t = (1-t)\mathbf{z}_1 + t\mathbf{z}_0, \quad t \in [0,1].
    \label{eq:cfm_path_en}
\end{equation}

The associated velocity along this path is defined as:
\begin{equation}
    \mathbf{u} = \mathbf{z}_0 - \mathbf{z}_1.
    \label{eq:cfm_velocity_en}
\end{equation}

In this setting, we parameterize the DiT  as a velocity network $v_\theta(\mathbf{z}_t, \mathbf{c}, t)$, which is trained to approximate the ground-truth velocity field. The learning objective is given by:
\begin{equation}
    \mathcal{L}_{\text{flow}} = \mathbb{E}_{t, \mathbf{z}_0, \mathbf{c}} \left[ \left\| v_\theta(\mathbf{z}_t, \mathbf{c}, t) - \mathbf{u} \right\|^2 \right],
    \label{eq:cfm_loss_en}
\end{equation}
where $\mathbf{c}$ denotes the conditioning information.



\section{Proposed method}
\label{sec:format}
\subsection{Overview}

The architecture of DMP-TTS is shown in Fig.~\ref{fig:model_architecture}(a). At its core are stacked DiT blocks operating on mel-spectrogram latents, trained to map Gaussian noise to target latent representations. Conditioning comes from three complementary inputs (content text, timbre, and style), which are provided by a text encoder, a speaker encoder, and a style encoder, respectively. Furthermore, as rhythm is central to speaking style, a duration predictor conditioned on text and style embeddings controls phoneme-level alignment. 

In the following sections, we detail the Style-CLAP, the chained CFG mechanism, and REPA.


\begin{table*}[t]
\centering
\caption{Comparison with baselines on params, capabilities, MOS, accuracy, speaker similarity, and WER. \textbf{Txt@Spk} and \textbf{Aud@Spk} denote, respectively, support for text- and audio-based style prompting under a specified target speaker timbre.}
\label{tab:comparison}
\setlength{\tabcolsep}{5pt}
\renewcommand{\arraystretch}{1.1}
\begin{tabular}{l c cc cc ccc c c}
\toprule
\multirow{2}{*}{\textbf{Method}} &
\multirow{2}{*}{\textbf{Params}} &
\multicolumn{2}{c}{\textbf{Capabilities}} &
\multicolumn{2}{c}{\textbf{MOS}$\uparrow$} &
\multicolumn{3}{c}{\textbf{Accuracy}$\uparrow$} &
\multirow{2}{*}{\textbf{Spk-Sim}$\uparrow$} &
\multirow{2}{*}{\textbf{WER}$\downarrow$} \\
\cmidrule(lr){3-4}\cmidrule(lr){5-6}\cmidrule(lr){7-9}
& & \textbf{Txt@Spk} & \textbf{Aud@Spk} &
\textbf{NMOS} & \textbf{QMOS} &
\textbf{Emotion} & \textbf{Energy} & \textbf{Rate} &
& \\
\midrule
GT             & -- & --     & --     & 3.86\textpm0.21 & 3.89\textpm0.09   & 0.68 & 1.00 & 1.00 & --   & 0.028 \\
GT-Recon       & -- & --     & --     & 3.74\textpm0.28 & 3.62\textpm0.12  & 0.62 & 0.80 & 0.97 & --   & 0.030 \\
CosyVoice \cite{du2024cosyvoice}      & 0.3B & \cmark & \xmark & 3.83\textpm0.26 & \underline{4.02\textpm0.13}   & 0.29 & 0.22 & 0.51 & 0.68 & 0.059 \\
CosyVoice2 \cite{du2024cosyvoice2}    & 0.5B & \cmark & \xmark & \underline{3.92\textpm0.22} & 3.95\textpm0.17    & 0.33 & 0.31 & 0.52 & \textbf{0.80} & 0.046\textsuperscript{*} \\
IndexTTS2\textsuperscript{\dag} \cite{zhou2025indextts2} & 1.0B & \xmark & \cmark & \textbf{4.03\textpm0.18} & \textbf{4.09\textpm0.13}   & 0.54 & 0.40 & 0.70 & \underline{0.76} & \textbf{0.028} \\
\midrule[0.8pt]
DMP-TTS (Audio)   & \multirow{2}{*}{0.3B} & \multirow{2}{*}{\cmark} & \multirow{2}{*}{\cmark} & 3.82\textpm0.23 & 3.83\textpm0.14   & \underline{0.55} & \underline{0.82} & \textbf{0.74} & 0.72 & 0.043 \\
DMP-TTS (Text)    &                      &                          &                          & 3.73\textpm0.27 & 3.77\textpm0.11   & \textbf{0.64} & \textbf{0.85} & \underline{0.73} & 0.71 & \underline{0.038} \\
\bottomrule
\end{tabular}

\vspace{-0.5\baselineskip}

\begin{flushleft}
\footnotesize \textsuperscript{\dag}\,IndexTTS2 provides textual control \emph{only} for emotion; therefore, we do not evaluate text-style prompting (Txt@Spk) and omit related metrics.\\
\footnotesize \textsuperscript{*}\,CosyVoice2 may concatenate the textual prompt into the synthesized output, inflating WER; we filter prompt text when computing WER.
\end{flushleft}
\vspace{-1.5\baselineskip}
\end{table*}

\subsection{Unified Multi-Modal Style Encoder}


To enable intuitive control over speech style, we build Style-CLAP, a unified style encoder (Fig.~\ref{fig:model_architecture}(b)) upon the pre-trained CLAP model. The pre-trained model inherently aligns text and audio in a shared embedding space, and we further fine-tune it for TTS style modeling.

We first curate textual style labels covering emotion, energy, and speech rate, while excluding descriptors unrelated to style (age, gender, pitch) to avoid overlap with speaker identity.
To refine the alignment between audio and text style embeddings, we fine-tune the model with an InfoNCE \cite{oord2018representation} loss. 
Specifically, the contrastive loss $\mathcal{L}_{\text{con}}$ is formulated as:

\begin{equation}
    \mathcal{L}_{\text{con}} = - \mathbb{E} \left[ \log \frac{\exp(\text{sim}(\mathbf{h}_a, \mathbf{h}_t)/\tau)}{\sum_{j=1}^{N} \exp(\text{sim}(\mathbf{h}_a, \mathbf{h}_{t,j})/\tau)} \right],
    \label{eq:infonce_en}
\end{equation}
where $\mathbf{h}_a = \mathcal{E}_a(\mathbf{A}_i)$ and $\mathbf{h}_t = \mathcal{E}_t(\mathbf{T}_i)$ are the embeddings extracted by the audio encoder $\mathcal{E}_a$ and text encoder $\mathcal{E}_t$, respectively. $\text{sim}(\cdot, \cdot)$ denotes cosine similarity, and $\tau$ is a temperature parameter.


However, a contrastive loss alone, while aligning modalities, does not guarantee that the learned representations are discriminative for specific style attributes. To address this, we incorporate multi-task supervision into the audio branch to enhance its representations. 
Specifically, we use cross-entropy loss $\mathcal{L}_{\text{ce}}$ for the discrete attribute—specifically, emotion—and mean squared error loss $\mathcal{L}_{\text{mse}}$ for continuous attributes, including speech rate and energy.
The overall training objective is a weighted sum of these components:
\begin{equation}
    \mathcal{L}_{\text{style}} = \mathcal{L}_{\text{con}} + \lambda_c \mathcal{L}_{\text{ce}} + \lambda_m \mathcal{L}_{\text{mse}} ,
\end{equation}
where $\lambda_c$ and $\lambda_m$ are balancing coefficients. This combined objective ensures the style encoder learns cross-modal and discriminative style representations.

\subsection{Chained Classifier-Free Guidance}
\label{sec:chained_cfg_en}
Classifier-Free Guidance \cite{ho2021classifier} enhances generation quality by learning with and without conditions through random condition dropping. Yet the standard \emph{all-or-nothing} setup yields only a global unconditional branch, limiting attribute disentanglement. As a result, the guidance scale strengthens content, timbre, and style \emph{together}, making it difficult to control the expressiveness of any single condition independently.


Inspired by MegaTTS 3 \cite{jiang2025megatts}, we introduce a chained classifier-free guidance (cCFG) mechanism for independent control over content, timbre, and style.
During training of the DiT, we adopt a hierarchical condition dropout strategy inspired by Vevo’s \cite{zhang2025vevo} definition of information levels—treating semantics (text) as high-level and acoustic attributes as progressively lower-level.
Concretely, we first drop the style condition $\mathbf{c}_{\text{style}}$ with probability $p_{\text{style}}$; if it is dropped, we then drop the timbre condition $\mathbf{c}_{\text{spk}}$ with probability $p_{\text{spk}}$; finally, we drop the text condition $\mathbf{c}_{\text{text}}$ with probability $p_{\text{text}}$ only when both style and timbre are absent.
In addition, we apply a \emph{style} perturbation during training—randomly feeding the speaker encoder with a different utterance from the same speaker to regularize the timbre branch and reduce style leakage.
At inference time, this training scheme enables chained guidance, where the final prediction $\hat{v}$ is given by:
\begin{equation}
\begin{aligned}
    \hat{v} = & \ v(\varnothing) 
            + s_{\text{text}} \left[ v(\mathbf{c}_{\text{text}}) - v(\varnothing) \right] \\
            & + s_{\text{spk}}  \left[ v(\mathbf{c}_{\text{text}}, \mathbf{c}_{\text{spk}}) - v(\mathbf{c}_{\text{text}}) \right] \\
            & + s_{\text{style}} \left[ v(\mathbf{c}_{\text{text}}, \mathbf{c}_{\text{spk}}, \mathbf{c}_{\text{style}}) - v(\mathbf{c}_{\text{text}}, \mathbf{c}_{\text{spk}}) \right],
\end{aligned}
\label{eq:chained_cfg_compact_en}
\end{equation}
where $v(\cdot)$ denotes the predicted velocity field. For clarity, we omit the sampling timestep $t$ and the intermediate latent state $\mathbf{z}_t$.
By tuning the guidance scales $s_{\text{text}}$, $s_{\text{spk}}$, and $s_{\text{style}}$, we can independently control the strength of each attribute in the generated speech.



\subsection{Representation Alignment}
\label{sec:repa_en}

To stabilize and accelerate the convergence of our multi-conditional TTS model, we adopt a Representation Alignment strategy. This method injects acoustic-semantic priors into the DiT backbone by incorporating knowledge from a pre-trained model. In particular, we utilize the audio encoder of Whisper \cite{radford2023robust} as a teacher to guide the learning of intermediate representations within the DiT.

Concretely, we take the final-layer output of the Whisper encoder as the teacher representation, $\mathbf{h}_{\text{whisper}} \in \mathbb{R}^{T_w \times D_w}$, and an intermediate-layer output of the DiT as the student representation, $\mathbf{h}_{\text{DiT}} \in \mathbb{R}^{T_d \times D_d}$. Since the two differ in sequence length and feature dimensionality, we first upsample $\mathbf{h}_{\text{DiT}}$ along the temporal axis and then apply a linear projection $\mathcal{P}$ to match the target dimension $D_w$. The alignment is achieved by minimizing a cosine similarity loss:
\begin{equation}
    \mathcal{L}_{\text{repa}} = 1 - \mathbb{E}_{t} \left[ \text{sim}(\mathcal{P}(\text{Upsample}(\mathbf{h}_{\text{DiT}}))_t, (\mathbf{h}_{\text{whisper}})_t) \right]
    \label{eq:repa_en}
\end{equation}
where $\text{sim}(\cdot, \cdot)$ denotes cosine similarity, and the expectation $\mathbb{E}_{t}$ is taken over the temporal dimension $t$.

\section{EXPERIMENTS}
\label{sec:exp}
\subsection{Dataset and evaluation}
We constructed an internal, high-quality Chinese speech dataset of approximately 300 hours, containing ~250k utterances from around 1,000 speakers. Emotion labels (happy, sad, angry, neutral, and fearful) were annotated using Qwen2.5-Omni \cite{xu2025qwen2}. To obtain accurate estimates of speech rate and energy, we employed the Silero VAD model\footnote{\url{https://github.com/snakers4/silero-vad}} to extract effective speech durations. Speech rate was estimated as the number of Chinese characters divided by the effective duration, and loudness was computed in LUFS using the \texttt{pyloudnorm} library. Finally, both energy and speech rate were categorized into three discrete levels. In addition, character-level timestamps were obtained using a forced alignment tool\footnote{\url{https://github.com/MahmoudAshraf97/ctc-forced-aligner}}.


For evaluation, we uniformly sampled 100 utterances from the training data to construct a test set with a balanced distribution of style attributes (emotion, energy, and speech rate); all style labels were manually verified by human annotators, and the synthesis texts were out-of-domain. We conducted \emph{cross-speaker} style transfer to assess disentanglement: each test utterance was randomly paired one-to-one with another utterance from a \emph{different} speaker, where one provided the timbre reference and the other supplied the content text and style. 
Word Error Rate (WER) was computed using the paraformer-zh \cite{gao2022paraformer} ASR model from FunASR. Speaker similarity was measured with Resemblyzer embeddings against the timbre-providing reference. Emotion accuracy was obtained using emotion2vec \cite{ma2024emotion2vec} on the synthesized audio and the style-providing reference. Energy and speech-rate accuracies follow the data-preparation pipeline, with both discretized into three levels. In addition, we conducted subjective listening tests and report NMOS (naturalness MOS) and QMOS (quality MOS).


\subsection{Implementation details}
\noindent\textbf{Mel-VAE.} Following a Kling–Foley \cite{wang2025kling} audio codec, 44.1 kHz waveforms are encoded by a mel encoder into 40-dimensional latents at 43 Hz, yielding an effective 1024× temporal downsampling.

\noindent\textbf{Style-CLAP.} 
We adopt the pre-trained CLAP\footnote{\url{https://huggingface.co/laion/clap-htsat-fused}}, inserting MLP auxiliary heads before the audio projector for style supervision.
Training is performed on 8$\times$A800 GPUs with a batch size of 128 for 50k steps, employing \emph{formant} perturbation to mitigate timbre leakage. The learning rate is $1\times10^{-5}$ with 5k warmup steps.


\noindent\textbf{TTS model.} The overall architecture follows the F5-TTS \cite{chen2024f5} base configuration, with DiT and text encoder depths and hidden dimensions aligned accordingly. The speaker encoder is initialized from the pre-trained cam++ \cite{wang2023cam++} model in CosyVoice. The duration predictor consists of Conv1D blocks followed by MLPs, we detach gradients from its inputs. Waveform synthesis is performed by a BigVGAN \cite{leebigvgan} vocoder. Following the recommendation in the REPA paper, we select the 6th DiT block as the student layer for REPA, and extract the target acoustic–semantic features using Whisper Large-v3. The TTS model is trained on 8$\times$A800 GPUs for 85k steps, with 38,400 frames per batch. The learning rate is $7.5\times10^{-5}$ with 20k warmup steps. For chained CFG, the dropout probabilities are set to 0.3 for style, 0.5 for timbre (conditional on style dropout), and 0.5 for text (conditional on both style and timbre dropout).


\subsection{Results analysis}
\label{sec:result}
\subsubsection{Comparison with Baselines}
We compare against large-scale open-source TTS baselines pretrained on $\sim$100k-hour corpora and evaluate them in zero-shot mode, whereas prior controllable-TTS systems (e.g., ControlSpeech) are English-only and lack Chinese releases, precluding direct comparison. As shown in Table~\ref{tab:comparison}, our method surpasses open-source baselines in \emph{style controllability}: with text prompts, emotion/energy/rate reach 0.64/0.85/0.73; with audio prompts, 0.55/0.82/0.74—exceeding the best baseline scores (0.54/0.40/0.70). For \emph{intelligibility}, we achieve WERs of 0.038 and 0.043 for text and audio prompts, respectively—second only to IndexTTS2 (0.028). In terms of perceptual quality, our NMOS/QMOS under audio prompting (3.82/3.83) are comparable to ground-truth recordings (3.86/3.89), effectively reaching real-speech–level naturalness. However, our \emph{speaker similarity} (0.71--0.72) remains below CosyVoice2 (0.80) and IndexTTS2 (0.76). We hypothesize that this gap arises from stronger generalization and speaker representations learned via large-scale pre-training in those models; additionally, information loss due to mel-spectrogram compression may attenuate speaker-discriminative cues.


Regarding prompt modality, text prompts yield more stable and higher style control, whereas audio prompts deliver higher naturalness (NMOS 3.82 vs.\ 3.73), suggesting richer prosodic and acoustic information in audio cues; this pattern also aligns with intuition—text provides discrete, interpretable control targets, while audio carries fine-grained prosody that enhances naturalness. 
Under the cross-speaker setting, we further observe comparable speaker similarity for audio- and text-conditioned synthesis, indicating that audio  style prompting does not leak timbre cues.


\begin{table}[t]
\centering
\caption{Ablation study on multi-task supervision (Sup.) and REPA.}
\label{tab:ablation}
\scriptsize
\setlength{\tabcolsep}{2.8pt}
\renewcommand{\arraystretch}{1.08}
\resizebox{0.95\columnwidth}{!}{%
\begin{tabular}{l ccc c c}
\toprule
\multirow{2}{*}{\textbf{Method}} &
\multicolumn{3}{c}{\textbf{Accuracy}$\uparrow$} &
\multirow{2}{*}{\textbf{Spk-Sim}$\uparrow$} &
\multirow{2}{*}{\textbf{WER}$\downarrow$} \\
\cmidrule(lr){2-4}
& \textbf{Emotion} & \textbf{Energy} & \textbf{Rate} & & \\
\midrule
DMP-TTS (Text)     & 0.64 & 0.85 & 0.73 & 0.71 & 0.038 \\
w/o Sup.        & 0.54 & 0.80 & 0.74 & 0.71 & 0.037 \\
w/o REPA        & 0.63 & 0.82 & 0.74 & 0.70 & 0.046 \\
\bottomrule
\end{tabular}
}
\vspace{-1.0\baselineskip}
\end{table}

\subsubsection{Ablation Study}
Our ablation study involves two comparisons: removing the multi-task supervision (Sup.) and disabling the REPA strategy. For all experiments, we used text-based prompts during inference. 

As shown in Table~\ref{tab:ablation}, removing Sup. notably weakens style control: Emotion accuracy drops from \textbf{0.64$\rightarrow$0.54} and Energy from \textbf{0.85$\rightarrow$0.80}, while Rate remains comparable (0.73 vs. 0.74); speaker similarity and WER show negligible change. In contrast, removing REPA primarily impacts linguistic fidelity: WER degrades from \textbf{0.038$\rightarrow$0.046}, with only minor changes in style accuracies and speaker similarity (0.71$\rightarrow$0.70). These results indicate that Sup. chiefly strengthens style control, whereas REPA aligns the DiT with Whisper’s acoustic–semantic space and lowers WER. Additionally, we observe that with REPA the model reaches intelligible speech at earlier training steps, indicating faster and more stable convergence.

\begin{figure}[t]
  \centering
  \subfigure[Speaker Similarity]{%
    \includegraphics[width=.49\linewidth]{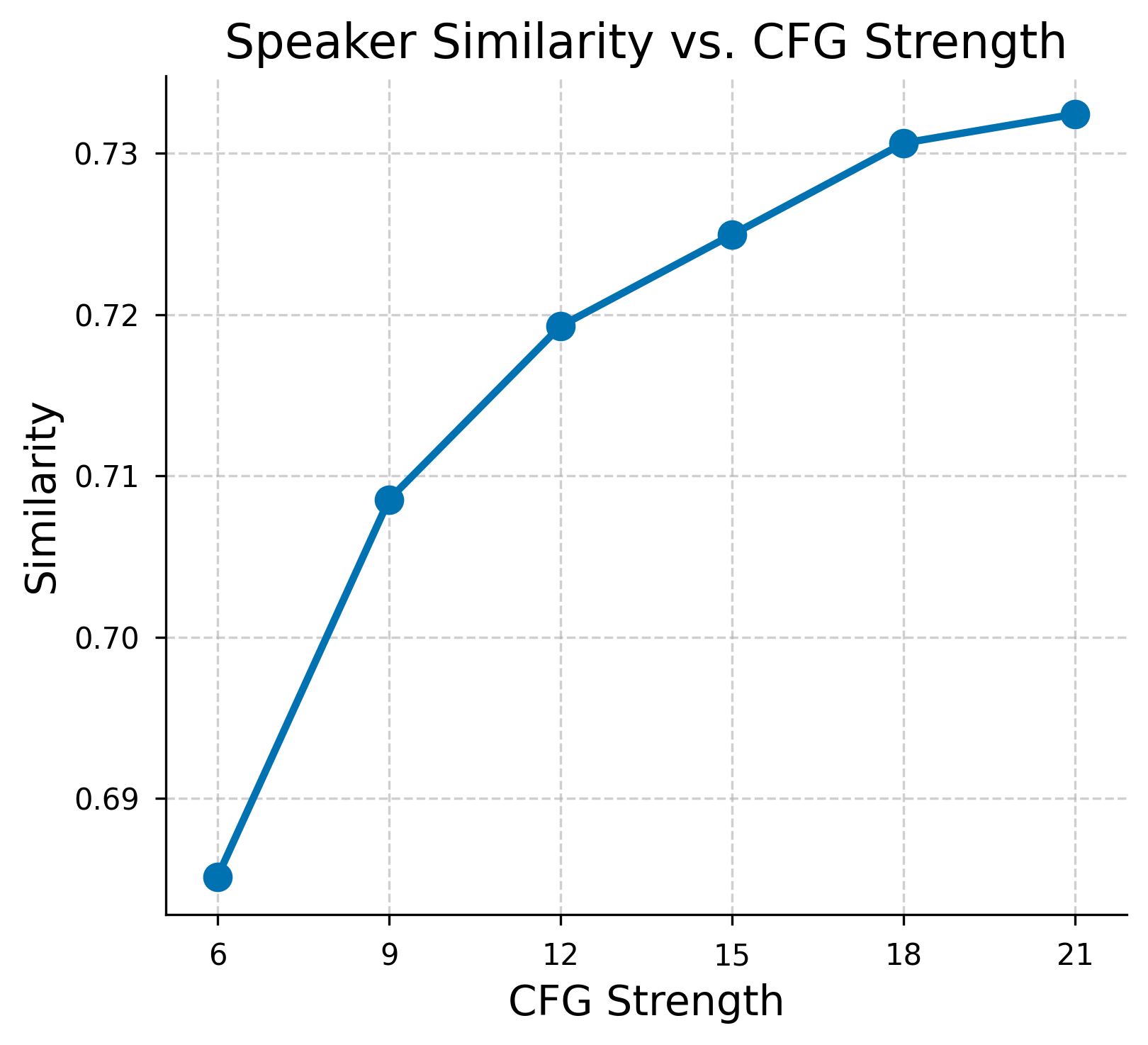}\label{fig:emo}}
  \hfill
  \subfigure[Emotion Accuracy]{%
    \includegraphics[width=.49\linewidth]{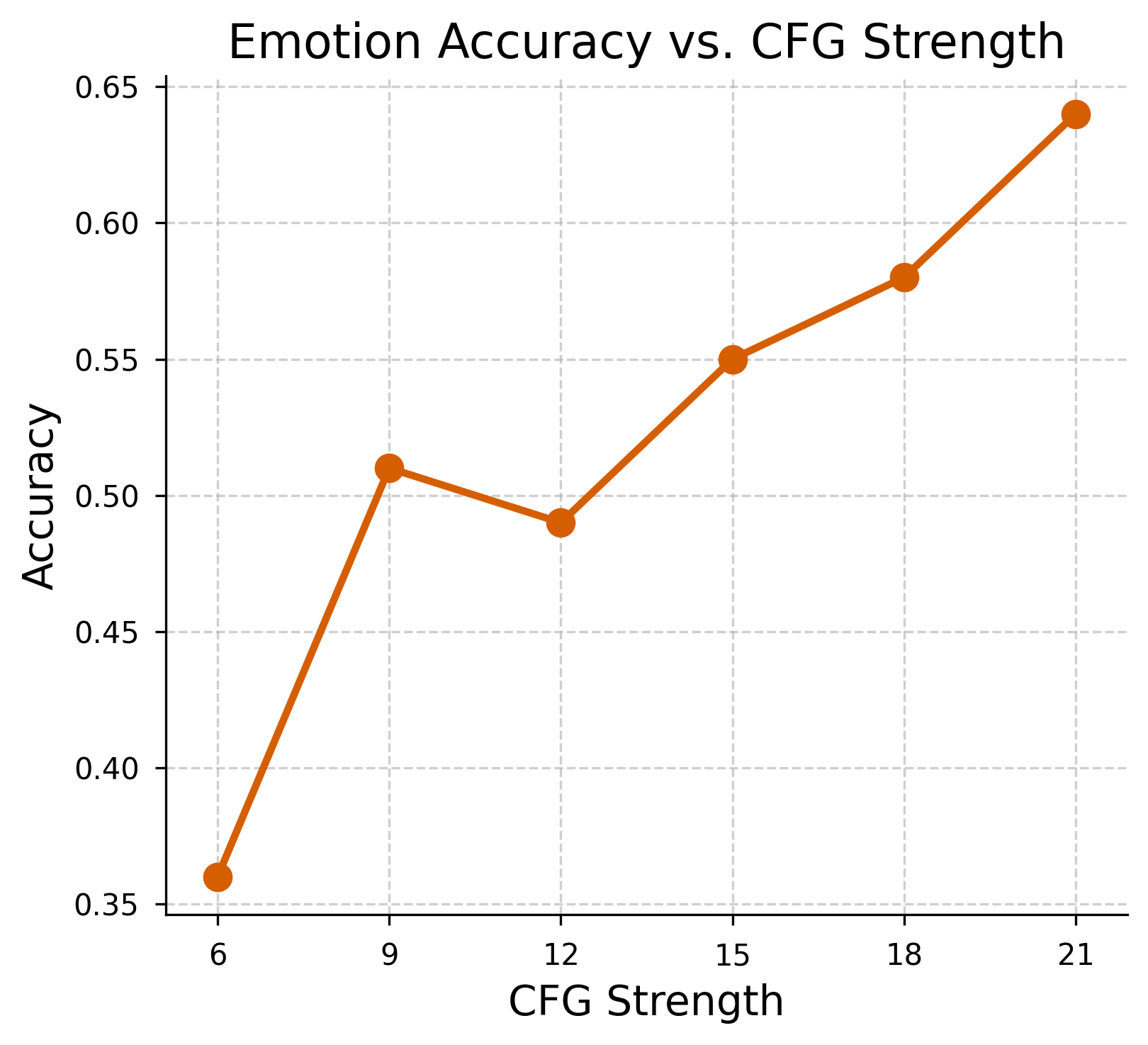}\label{fig:spk}}
  \vspace{-0.0\baselineskip}
  \caption{Effect of CFG strength on (a) speaker similarity and (b) emotion accuracy.}
  \vspace{-1.0\baselineskip}
  \label{fig:trend}
\end{figure}

\subsubsection{Effect of CFG Guidance Scales}
The CFG guidance scale governs the adherence to conditional inputs. 
We analyze its specific impact on both speaker similarity and emotion accuracy to understand the inherent relationship between these two attributes under varying guidance strengths.

Specifically, we analyze the impact of individually varying the
speaker or style guidance scale from 6.0 to 21.0 in steps of 3.0,
while holding the other scales fixed. As shown in Figure \ref{fig:trend}, as the
guidance scale increases, both speaker similarity (Figure \ref{fig:trend}(a)) and
emotion accuracy (Figure \ref{fig:trend}(b)) show a general upward trend. Subjective listening tests also indicate that a higher style guidance yields
more salient expressiveness (e.g., a “sad” tone becomes more pronounced), while a higher speaker guidance makes the synthesized voice sound closer to the target timbre. 
However, excessively high scales introduce over-conditioning that degrades naturalness and can slightly reduce the non-target attribute.
In practice, moderate scales strike a better balance between controllability and perceptual quality.


\subsection{Discussion}
Looking ahead, we outline two natural extensions of \sys. \textit{(i) Scaling style alignment and descriptions:} we plan to train the CLAP-based alignment on larger, multilingual corpora and to expand the style taxonomy with descriptions produced by Qwen2.5-Omni or stronger multi-modal LLMs, moving beyond static labels to dynamic style descriptors (e.g., energy trajectories) that capture temporal variation. \textit{(ii) Scaling the TTS backbone and enabling semi-supervised learning:} we aim to scale the main diffusion transformer to larger model sizes and datasets, and to support semi-supervised training in which utterances without textual style annotations are trained using only the CLAP audio branch as style embeddings, while maintaining disentanglement via contrastive and consistency objectives.

\section{Conclusion}
We presented \sys, a controllable TTS framework that disentangles speaker timbre and speaking style via a CLAP-based multi-modal style encoder and enables independent attribute adjustment through chained classifier-free guidance; a REPA strategy further stabilizes training and improves convergence. 
Experiments show that DMP-TTS delivers strong style controllability with both text and audio prompts, competitive intelligibility, and good naturalness. Ablations verify that multi-task supervision chiefly boosts style control, whereas REPA improves intelligibility and yields earlier intelligible speech. Our analysis of guidance scales shows that increasing guidance strength enhances the expressiveness of the targeted attribute (speaker timbre or speaking style).
We hope this work catalyzes advances in attribute disentanglement and training paradigms for controllable TTS. 

\vfill\pagebreak




{
\bibliographystyle{IEEEbib}
\bibliography{strings,refs}
}
\end{document}